\documentclass[prd,aps,A4,singlecolumn,superscriptaddress,preprintnumbers,nopacs,floatfix,amsmath,amssymb]{revtex4}
\newcommand{\lbl}[1]{\label{#1}}

\usepackage{bm}

\usepackage{amsmath}
\usepackage{graphicx}
\usepackage{amssymb}
\usepackage{mathrsfs}
\usepackage{bbm}

\newcommand{\be}{\begin{eqnarray}}
\newcommand{\ee}{\end{eqnarray}}

\begin{document}

%
%
%
\title{{\large\bf Re-dressing Emperor:  Four Dimensional 
Yang-Mills Theory, \\  Gauge Invariant Mass And   Fluctuating Three Branes
  }}
  
\author{Antti J. Niemi}
\email{Antti.Niemi@physics.uu.se}
\affiliation{Department of Physics and Astronomy, Uppsala University,
P.O. Box 803, S-75108, Uppsala, Sweden}
\affiliation{
Laboratoire de Mathematiques et Physique Theorique
CNRS UMR 6083, F\'ed\'eration Denis Poisson, Universit\'e de Tours,
Parc de Grandmont, F37200, Tours, France}
\author{Sergey Slizovskiy} 
\email{Sergey.Slizovskiy@physics.uu.se}
\affiliation{Department of Physics and Astronomy, Uppsala University,
P.O. Box 803, S-75108, Uppsala, Sweden}

\begin{abstract}
\noindent
We are interested in a gauge invariant  coupling between four dimensional Yang-Mills field and a three brane
that can fluctuate into higher dimensions. For this we interpret the Yang-Mills
theory as a higher dimensional bulk gravity theory with dynamics  that is governed by the 
Einstein action, and with a metric tensor constructed from the gauge field in a manner that
displays the original gauge symmetry as an isometry.
The brane  moves in this higher dimensional space-time under the influence of its bulk gravity,  with
dynamics  determined by the Nambu action. This  introduces the desired 
interaction between the brane and 
the gauge field in a way that preserves the original gauge invariance as an isometry of the induced metric.   
After a prudent change  of variables the result can be interpreted as  a gauge invariant and 
massive vector field  that propagates in the original space-time $\mathbb R^4$: The presence of the
brane becomes entirely invisible, except for the mass.
\end{abstract}



\maketitle

\vskip 1.0cm
\noindent
{\bf \large I: Introduction:}
\vskip 0.4cm
The existence  of a mass gap in four dimensional Yang-Mills theory remains unresolved \cite{Clay}.  
Here we propose a gauge invariant mass term that has its  origin
in higher dimensions.  For this we reformulate the Yang-Mills theory as a gravity theory with Einstein action in a higher 
dimensional space-time. This space-time  emerges when  we replace the matrix valued Lie algebra 
generators by Killing vector fields that act on an internal Riemannian manifold with an isometry group that coincides 
with the original gauge group. The standard $D=4$ flat space-time Yang-Mills action is obtained from
the higher dimensional  gravity action when we average  over the internal manifold. This computation of  the
averages over the internal manifold replaces the evaluation of the matrix traces over the Lie algebra generators in
the conventional formulation. We then proceed to introduce a three brane in this higher dimensional
space-time. Asymptotically the brane  stretches  into $\mathbb R^4$ but  it can locally  fluctuate into the internal manifold where it moves
under the influence of the bulk gravity and with  dynamics  determined by the Nambu action. 
The gravitational interaction of the brane leads to an effective interaction between the original Yang-Mills gauge field and the 
brane fluctuations. When viewed from the point  of view of the original flat four dimensional space-time, this can
be  interpreted in terms of  a massive vector field that resides in $\mathbb R^4$:
Much like in the conventional Higgs effect where a gauge field combines with a Higgs boson, the 
gauge field  now {\it entirely} eats up the higher dimensional brane fluctuations and becomes massive so that at the end,
there is nothing else left in the theory that reveals  the presence of a brane except the mass.
The internal Riemannian manifold can be chosen  to be any manifold with an isometry group that coincides with 
the original gauge group, and different choices give rise to different kind of mass terms. Examples include 
the group manifold itself and its  co-adjoint orbits.

\vskip 1.0cm
\noindent
{\bf \large II: Gauge Group as a Manifold:}
\vskip 0.4cm

The $SU(N)$ Yang-Mills action in $\mathbb R^4$ is
\begin{equation}
S_{YM} = - \frac{1}{2e^2}  \int d^4x   \, Tr \{ F_{\mu\nu} F_{\mu\nu} \} 
 =  \frac{1}{4e^2} \int d^4x \,   F^a_{\mu\nu} F^a_{\mu\nu} 
 \label{FF}
\end{equation}
with
\begin{equation}
F^a_{\mu\nu} (A) = \partial_\mu A^a_\nu - \partial_\nu A^a_\mu + f^{abc} A^b_\mu A^c_\nu
\label{F}
\end{equation}
The trace is over antihermitean matrices $T^a$ that represent the  Lie algebra of the gauge group $SU(N)$,
normalized so that
\[
Tr \{T^a T^b\} = - \frac{1}{2} \delta^{ab}
\]

Here we are interested in the interaction between the Yang-Mills field and a three brane. Such an interaction
can be  either difficult to introduce or then it lacks a proper interpretation  in the strictly four dimensional realm of (\ref{FF}).
For this we replace the matrices $T^a$ by Killing vector fields $K^a$ that act on an internal
Einstein manifold with  isometry group $SU(N)$. This
procedure is quite common in the context of Kaluza-Klein theories \cite{bengt} and 
nonlinear $\sigma$-models  \cite{hull}, but is rarely used in the conventional Yang-Mills theory.
Instead of the matrix trace we now have an integral over the internal manifold, and we shall assume  that this integral
generically gives us
\begin{equation}
Tr \{T^a T^b\}  \ \to \ \mu^{\dim [V_{int}]}  
\int \! \sqrt{g} \, d \vartheta \, g_{mn}(\vartheta) K^{am} (\vartheta) K^{bn}(\vartheta)  \ = \    \mu^{\dim [V_{int}]}  \cdot \frac{ \dim [ V_{int } ] }{\dim [SU(N)]}
\cdot V_{int} \,  \delta^{ab}
\label{int}
\end{equation}
Here $\vartheta^m$ are the  local
coordinates and $V_{int}$ is the volume of the internal manifold, and  $\mu$ is a mass scale. The
$K^{am}(\vartheta)$ are the components of Killing vectors
\[
K^a (\vartheta) = K^{am}\frac{\partial}{\partial\vartheta^m} 
\]
that satisfy the Lie algebra of the gauge group,
\begin{equation}
[ K^a , K^b ] \ = \  f^{abc} K^c
\label{Kcom}
\end{equation}
The metric is invariant,
\begin{equation}
\mathcal L^a g_{mn} \ = \  g_{mk}\partial_n K^{ak} + g_{kn} \partial_m K^{ak} + K^{ak} \partial_k g_{mn} \ = \ 0
\label{Lg}
\end{equation}
Furthermore, when
\begin{equation}
g_{mn}  = K^{a}_{m} K^{b}_{n} \,  \delta_{ab}
\label{gKK}
\end{equation}
we get from (\ref{Kcom}) the Maurer-Cartan equation
\[
\partial_m K^a_n - \partial_n K^a_m = f^{abc} K^b_m K^c_n
\]

In order to describe the interaction between the gauge field and the brane,   
instead of the Yang-Mills action (\ref{FF}) 
it is more convenient to take as the starting point the metric tensor
\begin{equation}
ds^2 = g_{\alpha \beta} dy^\alpha dy^\beta  
=  ( dx^\mu )^2 
+ g_{mn} (\vartheta) \{ d\vartheta^m +  K^{am} (\vartheta) A^a_\mu (x) dx^\mu \}
\{d\vartheta^n + K^{bn}(\vartheta)  A^b_\nu (x) dx^\nu \}
\lbl{ds}
\end{equation}
This metric tensor is akin the one that is widely employed in Kaluza-Klein theories \cite{bengt}. But here our goal
is very different.  We wish to interpret all of our results solely from the perspective of ordinary $\mathbb R^4$
Yang-Mills theory; Note that we have selected  $g_{mn}$ and $K^{am}$ to depend only on the internal coordinates
$\vartheta^m$, and  the gauge field $A_\mu^a$ depends only on the 
coordinates $x^\mu$ of $\mathbb R^4$. 

The relation (\ref{Lg})
implies that  the metric (\ref{ds}) remains intact under the following diffeomorphism
\begin{equation}
\vartheta^m \ \to \ \vartheta^m -  K^{am} (\vartheta) \varepsilon^a(x)
\label{coord1}
\end{equation}
provided
\begin{equation}
A^a_\mu (x) \ \to \ A^a_\mu (x) +  \partial_\mu  \varepsilon^a (x) + f^{abc} A^b_\mu (x) \varepsilon^c (x)
\label{cdr}
\end{equation}
This coincides with the familiar transformation law of a gauge field under infinitesimal gauge transformations.
To ensure that this diffeomorphism is an invariance in our reformulated Yang-Mills theory,
we choose the Einstein action in the higher dimensional space-time with a cosmological constant, and we 
evaluate it on the metric (\ref{ds}). This yields
\begin{equation}
S_E \ = \ \frac{1}{\kappa}  \! \int \! \sqrt{g} \, d^4x d\vartheta \, \{ R - 2\Lambda \} \ = \  \frac{1}{\kappa} \! \int \sqrt{g} \, d^4x d\vartheta \, \{
- \frac{1}{4} F^a_{\mu\nu} F^b_{\mu\nu} K^{a m} K^{b}_m + R_{int}  - 2\Lambda \}
\label{Ea}
\end{equation}
Here $R_{int}$ is the scalar curvature of the internal manifold. For  an Einstein manifold $R_{int}$   is a constant,
and we select the cosmological constant to cancel it. 

Notice that even though the rank of the metric  (\ref{gKK}) may
be smaller than the dimension of the gauge group, when we average over the internal manifold we may still 
obtain the result (\ref{int}).  In the following we shall tacitly assume this to be the case, and if we use 
the relation (\ref{int}) we are left with
\begin{equation}
S_E = \mu^{\dim[V_{int}]}   \cdot \frac{ \dim [ V_{int } ] }{\dim [SU(N)]} \, \frac {V_{int}} {4 \kappa } \int d^4x \, F^a_{\mu\nu} F^a_{\mu\nu}
\label{RF}
\end{equation}
This coincides with the original Yang-Mills action (\ref{FF}) with 
\[
e^2 =  \frac{\kappa}{  V_{int}} \mu^{-\dim [ V_{int}]} \, \cdot \frac{ \dim [ SU(N)  ] }{\dim [V_{int }]}
\]

\vskip 1.0cm
\noindent
{\bf \large III: Mass from Three-Brane:}
\vskip 0.4cm

We introduce a three brane $\mathcal B$ that is asymptotically  stretched into the 
space-time $\mathbb R^4$, but is allowed  to locally  fluctuate into the internal manifold \cite{sergey1}, \cite{sergey2}. 
This brane is described by 
\begin{equation}
\vartheta^m = X^m(x)
\label{X}
\end{equation}
and  we couple the brane to the Yang-Mills field by defining the brane dynamics to be determined by the Nambu action
\begin{equation}
S_{Nambu} = T \int d^4 x \sqrt{ G^{ind} }
\label{Gind}
\end{equation}
Here $G^{ind}$ is the determinant of the induced metric on the brane and $T$ is the brane tension.
We ensure that (\ref{Gind}) is finite by assuming that  the brane fluctuations are contractible
and have a compact support so that at large enough 
distances the world-sheet of the brane coincides with our space-time $\mathbb R^4$, 
\begin{equation}
X^m(x) \ =  \ 0 \ \ \ \ {\rm as} \ \ |x| \ > \ R
\label{infty}
\end{equation}
Here $R$ is some (finite) distance scale. This condition states
that for distances that are larger than $R$  the brane world-sheet merges  with the
original space-time $\mathbb R^4$, which is chosen so that it coincides with $\vartheta^m = 0$ in the
ambient space.

The induced brane metric is the  pull-back of the bulk 
metric  (\ref{ds}) to the world-sheet surface (\ref{X}), obtained by using the vielbein components
\[
{E_\mu}^\alpha \ = \ \delta_\mu^\alpha + \partial_\mu X^m \delta_m^\alpha
\]
Explicitely the result is
\[
G^{ind}_{\mu\nu} \ = \ E_\mu^\alpha g_{\alpha\beta} E_\nu^\beta = \delta_{\mu\nu} + \left( \partial_\mu X^m  + K^{am}A^a_\mu \right)
g_{mn} \left( \partial_\nu X^n  + K^{bn}A^b_\nu \right)
\]
The  determinant can be evaluated using Sylvester's theorem, 
\[
G^{ind} \ = \    1 + \delta^{\mu\nu} \left( \partial_\mu X^m  + K^{am}A^a_\mu \right)
g_{mn}  \left( \partial_\nu X^n + K^{bn}A^b_\nu \right) 
\]
and to the leading order in the brane fluctuation the Nambu action is
\begin{equation}
S_{Nambu} \ = \ T \int d^4x  + \frac{T}{2} \int d^4x \, \delta^{\mu\nu} J_\mu^m
g_{mn}(X) J_\nu^n \ + \ \dots
\label{Sexp}
\end{equation}
where we have defined 
\begin{equation}
J_\mu^m (X) \ = \   \partial_\mu X^m + K^{am} A^a_\mu 
\label{Jm}
\end{equation}
We remove the first term in  (\ref{Sexp}) by re-adjusting the cosmological constant in (\ref{Ea}). The second term
is a mass term:
\begin{equation}
S_{mass} \ = \ \frac{T}{2} \int d^4x \, g_{mn} (X)J_\mu^m
J_\mu^n 
\label{m}
\end{equation}
Due to the presence of the metric $g_{mn}(X)$  the mass apparently depends on the brane position but we shall soon find out
that  this is not the case, at least when the metric tensor admits the vielbein decomposition (\ref{gKK}).
But  if the rank of the metric tensor is smaller than the dimension of the gauge group the
number of massive components $J^m_\mu$ is smaller than the number of gauge fields, and in that case the massive
combinations in general may depend on the brane position.

We first verify that the Nambu action with our induced metric preserves the $SU(N)$ isometry (\ref{coord1}), (\ref{cdr}) of the
metric tensor (\ref{ds}),  corresponding to the original
gauge symmetry. For this we establish the gauge invariance of the current $J^m_\mu$:
We consider a diffeomorphism (\ref{coord1}) of the internal manifold, generated by the Killing vector $K^a$. We get
\begin{equation}
\delta_{\varepsilon}  (\partial_\mu X^m) \  = \ - K^{am} (X) \partial_\mu \varepsilon^a (x)
\label{bgt}
\end{equation}
while
\[
\delta_\varepsilon (K^{am} A^a_\mu) \  =  \  \delta_\varepsilon( K^{am} )\! \cdot \!  A^a_\mu + K^{am} \! \cdot \! \delta_\varepsilon (A^a_\mu) \ = \
(\varepsilon^b \mathcal L^b K^{am} )A^a_\mu + K^{am}(  \partial_\mu \varepsilon^a + 
f^{abc} A^b_\mu \varepsilon^c ) =  K^{am} \partial_\mu \varepsilon^a
\]
Consequently (\ref{Jm}) and in particular the Nambu action is gauge invariant {\it i.e.} we conclude that 
the $SU(N)$ isometry of (\ref{ds}) is preserved by the coupling between the gauge field and the brane.
 
Consider  next the quantity
\begin{equation}
B^a_\mu \ = \ K^a_m \partial_\mu X^m
\label{a}
\end{equation}
We compute
\[
\partial_\mu B^a_\nu - \partial_\nu B^a_\mu = \partial_\mu (\partial_\nu X^m K^a_m )
- \partial_\nu ( \partial_\mu X^m K^a_m ) \ = \    (\partial_n K^a_m - \partial_m K^a_n)  \partial_\nu X^m  \partial_\mu X^n
\]
\[ 
= \ - f^{abc} K^b_n K^c_m \partial_\nu X^m \partial_\mu X^n \ = \
 - f^{abc} B^b_\mu B^c_\nu
\]
Consequently (\ref{a}) obeys the Maurer-Cartan equation {\it i.e.} it is a pure gauge. In particular we can write 
\begin{equation}
B_\mu \ \equiv \ B^a_\mu T^a \ =  \  \partial_\mu \mathcal U^{-1} \!\!  \cdot \mathcal U
\label{BR}
\end{equation}
where $T^a$ are matrices in a defining representation of the gauge group $SU(N)$ and $\mathcal U$ is an element of the gauge group.
We introduce the vielbein basis
\begin{equation}
\hat e^a \ \equiv \ e^a_{i} T^i  \ = \  \mathcal U \,  T^a \, \mathcal U^{-1} 
\label{eU}
\end{equation}
Next we introduce the composite vector field
\begin{equation}
\mathcal J^i_{\mu}\, T^i  \ = \  (A^a_\mu + B^a_\mu )\,  e^a_i T^i 
\ = \ \mathcal U ( \partial_\mu + A_\mu ) \mathcal U^{-1}
\label{suJ}
\end{equation}
This vector is diffeomorphism {\it a.k.a.} gauge invariant under (\ref{coord1}), (\ref{cdr}).  When
we resolve (\ref{suJ}) for $A^a_\mu$ and substitute the result in (\ref{F}) we get
\[
F^a_{\mu\nu}(A) = \left( \partial_\mu \mathcal J^i_{\nu } - \partial_\nu \mathcal J^i_{\mu } + f_{ijk} \mathcal J^j_{\mu} \mathcal
J^k_{\nu } \right)  e^a_i \ \equiv \ F^i_{\mu\nu }  (\mathcal J) e^a_i
\]
Furthermore, when we {\it assume} that the metric tensor has the vielbein decomposition
(\ref{gKK}) we can also write the Nambu (mass) term entirely in terms of (\ref{suJ}). Combining the
Nambu action with the Yang-Mills action we then get
the following manifestly diffeomorphism {\it a.k.a.} gauge invariant action 
\[
S_{YM} + S_{Nambu} \ = \  \int d^4x \,  \left\{   \frac{1}{4e^2} F^i_{\mu\nu}(\mathcal J)F^i_{\mu\nu} (\mathcal J) \ + \  T \sqrt{ 1 +
 \mathcal J^i_{\mu }  \mathcal J^i_{\mu } } \right\} 
 \]
 \begin{equation}
= \  \int d^4x \,  \left\{   \frac{1}{4e^2} F^i_{\mu\nu}(\mathcal J)F^i_{\mu\nu} (\mathcal J) 
 \ + \  \frac{T}{2} 
\mathcal J^i_{\mu }  \mathcal J^i_{\mu } + \dots \right\}
\label{mass}
\end{equation}
Intrinsically this action describes the interaction between the Yang-Mills field $A^a_\mu$ with the three brane that fluctuates into
the internal manifold. But remarkably, when we write it in terms of the variable $\mathcal J^i_\mu$, it depends {\it only} on this
variable and {\it all} reference to higher
dimensions and in particular to the fluctuating  brane has disappeared:  The action (\ref{mass}) has a {\it direct} interpretation in terms of a  
massive vector field with $SU(N)$ invariant dynamics that takes place in the {\it original} space-time $\mathbb R^4$. We can also 
interpret this so that the  gauge field has "eaten up" the brane fluctuations and the result is the massive vector field $\mathcal J^i_\mu$,
furthermore with translationally invariant dynamics in the original flat space-time $\mathbb R^4$ since 
all dependence on the brane position has also disappeared: The {\it only } thing that reveals the presence of the brane in our
final theory is the presence of the mass term in $\mathbb R^4$.

More generally, we can show that the Nambu action is both $SU(N)$ isometric and independent of the brane position whenever the 
Killing vectors act transitively on the internal manifold, and the dimension of the internal manifold does not exceed
the number of the Killing vectors. This follows  directly from the previous construction: We have verified that the vector field
(\ref{Jm}) is gauge invariant {\it i.e.} its Lie derivative along the flow (\ref{coord1}), (\ref{cdr}) vanishes. Consequently we can locally
introduce a diffeomorphism  generated by the Killing vectors that  brings the brane coordinates to a constant value, for example
\[
X^m(x) = 0
\]
Consequently we can write the mass term patch-wise as
\[
S_{mass} \ = \ \frac{T}{2} \int \! d^4x \,  g_{mn}(0) J^m_\mu J^n_\nu
\]
which establishes the independence on the brane position.

\vskip 1.0cm
\noindent
{\bf \large IV: Anomalies and Monopoles:}
\vskip 0.4cm

Since  the variable $\mathcal J_\mu^i$ in (\ref{suJ}) is gauge invariant,  {\it any} Lorentz invariant action constructed 
from  it is also gauge invariant. But in order to motivate the introduction of natural candidates we  re-introduce
the brane variable and re-write the mass contribution to the Nambu action 
in the following  standard (Skyrme) form of a gauged non-linear $\sigma$-model,
\[
S_{mass} \ = \ - \frac{T}{4} \int d^4x \, Tr \{ \mathcal U(\partial_\mu + A_\mu )\mathcal U^{-1} \cdot \mathcal U(\partial_\mu + A_\mu )\mathcal U^{-1} 
\}
\ = \  \frac{T}{2} \int d^4x \, g_{mn}(X) \nabla_\mu X^m \nabla_\mu X^n
\]
This allows us to better relate our construction to known results   \cite{hull}, \cite{witten}, \cite{faddeev}. The
covariant derivative is defined by
\begin{equation}
\nabla_\mu X^m \ =  \ \partial_\mu X^m + K^{am}A^a_\mu
\label{covder}
\end{equation}
This $\sigma$-model version proposes us to  consider additional terms that have a natural $\sigma$-model 
interpretation. A general class of such terms is obtained
by starting from the four-form \cite{hull}
\begin{equation}
\int_{\mathcal B}  K_{mnpq} (\vartheta) d\vartheta^m  d\vartheta^n d\vartheta^p d\vartheta^q
\label{B1}
\end{equation}
Here the integral extends over the entire four dimensional world-sheet of the fluctuating  three brane. 
When we pull-back (\ref{B1})  into $\mathbb R^4$ and replace derivatives  with covariant derivatives 
we obtain a diffeomorphism invariant {\it a.k.a.} gauge invariant action in $\mathbb R^4$
under (\ref{coord1}), (\ref{cdr}) 
\begin{equation}
S_{K} \ = \ \frac{1}{4!} \int d^4x \,  \epsilon^{\mu\nu\rho\sigma} K_{mnpq}(X)    \nabla_\mu X^m \nabla_\nu
X^n \nabla_\rho X^p \nabla_\sigma X^q
\label{SK}
\end{equation}
provided
\[
\mathcal L^a K \ = \ 0
\]
Furthermore, since the brane fluctuation has a compact support we can interpret its world-sheet to be  the  boundary 
of a contractible five dimensional disk $\mathcal D_5$ in the internal manifold. 
Due to the boundary condition that at large distances the brane
coincides with $\mathbb R^4$ the disk  also  includes the point $\vartheta^m = 0$, 
which corresponds to the brane position of the original space-time $\mathbb R^4$.  We then use Stokes theorem to 
convert the integral  (\ref{B1})  into
an integral over the entire disk $\mathcal D_5$. The result is an integral of the form
\begin{equation}
\int_{\mathcal D_5} H_{mnpqr} (\vartheta) d \vartheta^m d \vartheta^n d\vartheta^p d\vartheta^q d\vartheta^r
 \label{D5}
 \end{equation}
 where $ H_{mnpqr} (\vartheta) $ are the components of the closed five-form,
 \[
  H_{mnpqr}(\vartheta)  d \vartheta^m d \vartheta^n d\vartheta^p d\vartheta^q d\vartheta^r  \ = \ d  \{ K_{mnpq}(\vartheta)   d \vartheta^m d \vartheta^n d\vartheta^p d\vartheta^q\}
 \]
However,  if we allow the five-form $H$ in (\ref{D5}) to be closed but not exact  the ensuing four-form $K$ can only be introduced locally. In that case
the extension of (\ref{D5})  into a diffeomorphism invariant quantity can not be constructed simply by minimal substitution.
An example is the following Wess-Zumino functional  \cite{witten}, \cite{faddeev}, \cite{hull},
\[ 
S_{WZ} \ = \ - \frac{i}{ 2 \pi^2 \cdot 5 ! }\int_{\mathcal D_5} d^5x \, \epsilon^{\alpha\beta\gamma\delta\eta} Tr ( B_\alpha B_\beta B_\gamma B_\delta B_\eta ) 
\]
This corresponds to the closed five-form
\[
H_{mnpqr} (\vartheta) \ = \  K_m^a K_n^b K_p^c K_q^d K_r^e \cdot Tr [ T^a T^b T^c T^d T^e ]
\]
Its diffeomorphism invariant extension is
\[
S_{WZ} \ = \ - \frac{i}{4! \cdot 2\pi^2} \int d^4 x \, \epsilon^{\mu\nu\rho\sigma} \left[ d_{abc} A^a_\mu \partial_\nu A^b_\rho B^c_\sigma \right.
\]
\[
\left. + C_{abcd}
\left( A^a_\mu A^b_\nu A^c_\rho B^d_\sigma - \frac{1}{2} A^a_\mu B^b_\nu A^c_\rho B^d_\sigma 
- A^a_\mu B^b_\nu B^c_\rho B^d_\sigma \right) + K_{mnpq} \partial_\mu X^m  \partial_\nu X^n \partial_\rho X^p \partial_\sigma X^q \right]
\]
where
\[
\frac{1}{2} d_{abc} = Tr [ T^a \{ T^b , T^b\} ] \ \ \ \ \& \ \ \ \ \  C_{abcd} \ = \ Tr[ T^a T^b T^c T^d]
\]
This is invariant under (\ref{bgt}) only if \cite{faddeev}
\begin{equation}
\frac{ \delta S_{WZ} }{ \delta \epsilon^a (x) } \ =  \  \frac{i}{24\pi^2} \epsilon^{\mu\nu\rho\sigma} \partial_\mu \left(
d_{abc} A^b_\nu \partial_\rho A^c_\sigma + C_{abcd} A^b_\nu A^c_\rho A^d_\sigma \right) \ = \ 0
\label{ano}
\end{equation}
When (\ref{ano}) is non-vanishing we have the familiar non-Abelian anomaly equation \cite{faddeev}, due to a single Weyl fermion
in interaction with the gauge field. We can interpret this in alternative ways:  The presence of
a gauge anomaly in a Yang-Mills theory with Weyl fermions leads to a breaking of 
diffeomorphism invariance in space-time fluctuations away from $\mathbb R^4$. 
Alternatively, a non-Abelian gauge  anomaly that arises from  Weyl fermions
can  be removed by allowing for appropriate three brane fluctuations that cancel those that emerge from the
Weyl fermions. In this manner our approach provides a very natural interpretation and setting for the consistent quantization
of anomalous gauge theories  \cite{faddeev}.  

We also note that in the present approach we entirely avoid the coventional introduction of an {\it ad hoc} five-dimensional disk \cite{hull},
\cite{witten}, \cite{faddeev}. This disk
has now a natural geometric interpretation in the context of our higher dimensional ambient space, and its boundary is the fluctuating three brane.

We return to (\ref{eU}).  We choose $H^\alpha$ to be the Cartan subset of the $SU(N)$ generators $T^a$ and we introduce the
ensuing subset $m^\alpha_i$ of the vielbeins  $e^a_i$ in (\ref{eU}) \cite{fad}, 
\begin{equation}
\hat m^\alpha \ = \ m^\alpha_i T^i \ = \ \mathcal U H^\alpha \mathcal U^{-1}
\label{ma}
\end{equation} 
It is straightforward to verify that
\begin{eqnarray*}
[ \hat m^\alpha , \hat m^\beta ] \ & = & \ 0 \\
\{ \hat m^\alpha ,\hat  m^\beta \} \ & = & \ d^{\alpha\beta\gamma}\hat m^\gamma \\
Tr( \hat m^\alpha \partial_\mu \hat m^\beta) \ & = & \ 0
\end{eqnarray*}
Using (\ref{BR}) we can also show that
\[
d\hat m^\alpha \ = \ [ \hat m^\alpha , B ]
\]
We introduce the following closed two-forms,
\begin{equation}
\Omega_H^\alpha \ = \ Tr ( \, H^\alpha [ \, \mathcal U^{-1} d \mathcal U \, , \, \mathcal U^{-1} d \mathcal U \, ] ) \ 
= \ f_{ijk} \, m_i^\alpha \partial_\mu m_j^\beta  \partial_\nu
m_k^\beta  \, dx^\mu \wedge dx^\nu
\label{Omega}
\end{equation}
These are the symplectic two-forms on the orbit $SU(N)/U(1)^{N-1} .$ Recall that according to the Borel-Weil theorem each of the 
linear combinations
\[
\sum_\alpha n_\alpha \Omega_H^\alpha
\]
where $n_\alpha \in \mathbb Z$ corresponds to an irreducible representation of $SU(N)$. We can show that
\begin{equation}
B_\mu dx^\mu \ = \ \mathcal U  \partial_\mu \mathcal U^{-1} dx^\mu  \ = \ C_H  \cdot \hat m + [ d\hat m , \hat m ] 
\label{Bmon}
\end{equation}
where 
\begin{equation}
\partial_\mu C_{H \, \nu}^\alpha   - \partial_\nu C_{H \, \mu}^\alpha \ = \ \Omega_{H \, \mu \nu}^\alpha
\label{C}
\end{equation}
This  reveals  a relation between the $SU(N)$ magnetic monopoles in the original space-time
$\mathbb R^4$, representations of $SU(N)$,
and the non-triviality of the  topological structure of the three brane $\mathcal B$.

\vskip 1.0cm
\noindent
{\bf \large V: SU(2) as an Example:}
\vskip 0.4cm

As an explicit example we consider  the case of $SU(2)$. For the internal manifold we first take  $SU(2) \sim \mathbb S^3$. 
We use the following explicit Euler angle parametrization
\begin{equation}
\mathcal U = -i \left( \begin{matrix} \sin \frac{\theta}{2} e^{\frac{i}{2}\phi_+} &
- \cos \frac{\theta}{2} e^{ \frac{i}{2} \phi_-} \\
- \cos \frac{\theta}{2} e^{- \frac{i}{2} \phi_-} & - \sin \frac{\theta}{2} e^{-\frac{i}{2}
\phi_+} \end{matrix} \right)
\lbl{UEu}
\end{equation}
where  $0 \leq \theta \leq \pi$ and
$0 \leq \phi_{\pm} \leq 2\pi$ are local coordinates on $\mathbb S^3$.
The natural metric $g_{mn}$ ($m,n = 1,2,3$)  on $\mathbb S^3$ is
the bi-invariant Killing two-form,
\begin{equation}
ds^2 = 2 \, Tr( d\mathcal U d\mathcal U^{-1} ) = g_{mn} d\vartheta^m d\vartheta^n 
=  (d \theta)^2 + \sin^2\!  \frac{\theta}{2} \,  (d\phi_+)^2 + \cos^2 \! \frac{\theta}{2}  \, (d\phi_-)^2
\label{bim}
\end{equation}
We write the Maurer-Cartan one-form as follows,
\begin{equation}
B_\mu \ = \ \mathcal U d \mathcal U^{-1}  = {B^a_{m}} d\vartheta^m \frac{1}{2i} \tau^a
\lbl{mc2}
\end{equation} 
where $\tau^a$ are the Pauli matrices. We relate 
the components ${B^a_{m}}$  to the dreibeins for the metric (\ref{bim}),
\begin{equation}
g_{mn} = \delta_{ab} {B^a_{m}}  {B^b_{n}} 
\lbl{gGG}
\end{equation}
The one-forms ${K^a}  = {B^a_m} d\vartheta^m$ are subject to the $SU(2)$ Maurer-Cartan equation
\begin{equation}
d{B^a}  = - \frac{1}{2} \epsilon^{abc} {B^b}  \wedge {B^c} 
\lbl{dg}
\end{equation}
Explicitely, we write
\begin{eqnarray}
{B^1}   & = & {n}^1  d \psi_+ -  {e_2}^1 \, d\theta 
\lbl{lhmc1}
\\
{B^2}  & = & {n}^2   d\psi_+ -  {e_2}^2  \, d\theta 
\\
{B^3}  & = &  {n}^3 d\psi_+ - \  d\psi_-
\lbl{lhmc3}
\end{eqnarray}
where we have defined
\begin{equation*}
\psi_{\pm} = \frac{1}{2} (\phi_+ \pm \phi_-) 
\end{equation*}
and we have introduced the right handed unit triplet
\begin{equation}
\vec e_1 \ = \ \left( \begin{matrix} \cos \psi_- \cos \theta \\ \sin \psi_- \cos\theta \\ - \sin\theta \end{matrix} \right) \ \ \ \ \ \  \& \ \ \ \ \ 
\vec e_2 \ = \ \left( \begin{matrix} - \sin \psi_-   \\ \cos \psi_-  \\ 0 \end{matrix} \right) 
\ \ \ \& \ \ \  \ \  \vec n \ = \    \left( \begin{matrix} \cos\psi_- \sin \theta 
\\ \sin \psi_- \sin\theta \\  \cos\theta \end{matrix} \right) 
\lbl{e1e2n}
\end{equation}
There are  the three invariant Killing vector fields 
\[
K^a = (K^a)^m \frac{ \partial}{\partial \vartheta^m} \ \ \ \ \ (m=1,2,3)
\]
that can be identified as the canonical duals of the one-forms $B^a$.  
With (\ref{lhmc1})-(\ref{lhmc3}) this gives us the explicit realization
\begin{equation}
K^1  = \left\{ \sin \psi_- \partial_\theta +  \cos\psi_- \cot \theta 
\partial_{\psi_-} \right\} \ + \ \frac{\cos\psi_-}{\sin\theta} \partial_{\psi_+} 
\lbl{KT1}
\end{equation}
\begin{equation}
K^2 
= \left\{ - \cos \psi_- \partial_\theta + \sin\psi_- \cot \theta \partial_{\psi_-} \right\}  \ 
+ \ \frac{\sin\psi_-}{\sin\theta} \partial_{\psi_+} 
\lbl{KT2}
\end{equation}
\begin{equation}
K^3   =  -  \partial_{\psi_-} 
\lbl{KT3}
\end{equation}
and the commutators of the Killing vectors determine a representation
of the $SU(2)$ Lie algebra, 
\begin{equation}
[ K^a , K^b ] = - \epsilon^{abc} K^c
\lbl{KKK}
\end{equation}
Using (\ref{Bmon}), (\ref{e1e2n})  we write
\begin{equation}
B_\mu \ = \ \mathfrak C_\mu \hat n  \ + \ [ \hat n , \partial_\mu \hat n]
\label{Bknot}
\end{equation}
where
\[
\mathfrak C_\mu \ = \  \vec e^{\, +} \cdot  \partial_\mu \vec e^{\, -} 
\]
with
\[
\vec e^\pm \ = \ \frac{1}{2} e^{-\psi_+} ( \vec e_1 \pm i \vec e_2 )
\]
Explicitely, in terms of the angular variables in (\ref{e1e2n}) 
\begin{equation}
\mathfrak C_\mu = - \frac{1}{2} ( \cos \theta \, \partial_\mu \psi_- + \partial_\mu \psi_+ )
\label{aknot}
\end{equation}
and for (\ref{C}) we get
\[
\partial_\mu \mathfrak C_\nu - \partial_\nu \mathfrak C_\mu \ = \ \vec n \cdot \partial_\mu \vec n \times \partial_\nu \vec n
+  \Sigma_{\mu\nu} 
\]
where 
\[
\Sigma_{\mu\nu} \ = \ - \frac{1}{2} [ \partial_\mu , \partial_\nu ] \psi_+
\]
is the familiar Dirac string tensor. Indeed, in (\ref{aknot}) we dentify the familiar structure of pointlike Dirac monopoles. We note
that this  structure is also intimately related to the presence of knot-like configurations in the space $\mathbb R^3$ \cite{lud}, and
these knots are the natural candidates for describing the (glueball) spectrum of the Yang-Mills theory.

Instead of $\mathbb S^3$, we can also take the internal manifold to be the co-adjoint orbit
$SU(2)/U(1) \sim \mathbb S^2$. The Killing vectors are now
\begin{equation}
K^1  = - \sin \phi \, \partial_\theta - \cot \theta \sin \phi \, \partial_\phi
\label{k1}
\end{equation}
\begin{equation}
K^2
= \cos \phi \, \partial_\theta - \cot \theta \sin \phi \, \partial_\phi 
\label{k2}
\end{equation}
\begin{equation}
K^3   =  \partial_\phi
\label{k3}
\end{equation}
The rank of the metric tensor on $\mathbb S^2$ is two, but for the integral (\ref{int}) we get
\begin{equation}
 \int\limits_{\mathbb S^2}  \! \sin\theta \, d \theta d\phi \, g_{mn} K^{am}  K^{bn}  \ = \   4 \pi \cdot \frac{2}{3}  \delta^{ab}
\label{int2}
\end{equation}
and consequently we obtain the Yang-Mills action from (\ref{Ea}), (\ref{RF}). But (\ref{m}) now gives a non-vanishing
mass to only two of the vector fields $J^m_\mu$, and the massless combination is
\[
J_\mu \ = \ \sin\phi(x)  \sin \theta(x) \, A^1_\mu + \cos\phi(x) \sin \theta(x) \, A^2_\mu + \cos \theta(x) \, A^3_\mu
\]
where $\phi(x)$ and $\theta(x)$ are the spherical coordinates of the brane position; We note that by properly
implementing the diffeomorphisms (\ref{coord1}), (\ref{cdr}) we can locally transport the brane position {\it e.g.}
to the north-pole $\theta = 0$ so that the massless mode becomes $A_\mu^3$.  This corresponds
to selecting a unitary gauge in the original Yang-Mills theory.

\vskip 1.0cm
\noindent
{\bf \large VI: Conclusions:}
\vskip 0.4cm

In conclusion, we have  considered a gauge invariant 
coupling between a four dimensional $SU(N)$ gauge field and a three brane.
For this we have reformulated the four dimensional Yang-Mills theory as a gravity theory in a higher
dimensional space-time, by   replacing the Lie algebra generators with Killing vector fields. 
This enables us to employ the Nambu action to introduce  an interaction between the Yang-Mills field and the 
three brane that fluctuates in the ensuing higher dimensional space-time. We have shown that  the Nambu action 
for the brane leads to a vector field mass.
The final theory describes the interactive dynamics of massive and massless vector fields in the original flat Euclidean four-space,
with the mass constituent depending on the choice of the internal manifold. In particular, all other reference to the three brane besides
the presence of a mass term in $\mathbb R^4$ becomes
entirely removed. We have also investigated topologically nontrivial brane structures, and extablished their connection
to magnetic monopoles in $\mathbb R^4$.
Our results suggest that the mass gap in 
Yang-Mills theory could well have its origin in higher dimensions. Or, at least our construction appears to give a novel and 
good motivation for introducing  certain familiar Skyrme-like effective fields to describe both the mass gap and the gauge anomaly 
in a continuum Yang-Mills theory. From the four dimensional point of view these theories are non-renormalizable. 
However, with the present higher dimensional gravity/membrane formulation,  maybe there is a  completion into a 
consistent theory. 

\vskip 0.2cm

We thank M. Chernodub for discussions. This work has been partially supported by a grant from Vetenskapsr\.adet.

\end{document}